\documentclass[11pt]{article}
\usepackage{graphicx}
\usepackage{setspace}
\usepackage{color}
\usepackage[super,sort&compress]{natbib}
\usepackage{multirow}
\usepackage{bm}
\usepackage{amsfonts}
\usepackage{amssymb}
\usepackage{enumerate,ulem}
\usepackage{float}
\usepackage{ulem}
\usepackage{color}

\definecolor{darkgreen}{RGB}{32,127,16}

\begin{document}

\begin{center}
{\bf \large Onsager's Wien Effect on a Lattice}
\end{center}

\begin{center}
V. Kaiser$^{1,2,*}$, S. T. Bramwell$^3$, P. C. W. Holdsworth$^2$, R. Moessner$^1$
\end{center}

\noindent
{\small
{\it 1. Max-Planck-Institut f\"ur Physik komplexer Systeme, N\"othnitzer Stra{\ss}e 38, 01187 Dresden,  Germany.}\\
{\it 2. Universit\'e de Lyon, Laboratoire de Physique, \'Ecole Normale Sup\'erieure de Lyon, 46 all\'ee d'Italie,
69364 Lyon CEDEX 07, France.}\\
{\it 3. London Centre for Nanotechnology and Department of Physics and Astronomy, University College London, 17--19 Gordon Street, London, WC1H OAH, UK.}}

\vspace{0.5cm}

{\bf 
The Second Wien Effect describes the non-linear, non-equilibrium response of a 
{weak} electrolyte in {moderate to high} electric fields. Onsager's 
1934 electrodiffusion theory~\cite{Onsager} along with various extensions~\cite{Braun,Noolandi} 
has been invoked for systems and phenomena as diverse as 
solar cells~\cite{Koster,Yuan},
surfactant solutions~\cite{Ran},
water splitting reactions~\cite{Strathmann, Mathe},
dielectric liquids~\cite{Park},
electrohydrodynamic flow~\cite{Ryu},
water and ice physics~\cite{Eigen}, 
electrical double layers~\cite{Bruesch},
non-Ohmic conduction in semiconductors~\cite{Pai} and oxide glasses~\cite{Tomoz},
biochemical nerve response~\cite{Tsong}
and magnetic monopoles in spin ice~\cite{BramwellGiblin}. In view of this technological importance and the experimental ubiquity of such phenomena, it is surprising that Onsager's Wien effect has never been studied by numerical simulation.
Here we present simulations of a lattice Coulomb gas, treating the widely applicable case of a double equilibrium for free charge generation. We obtain detailed characterisation of the Wien effect and confirm the accuracy of the analytical theories as regards the field evolution of 
the free charge density and correlations. We also demonstrate that simulations can uncover further corrections, such as how the field-dependent conductivity may be influenced by details of microscopic dynamics. We conclude that lattice simulation offers a powerful means by which to model and to investigate system-specific corrections to the Onsager theory, and thus constitutes a valuable tool for detailed theoretical studies of the numerous practical applications of the Second Wien Effect. 
}

\newpage

Onsager's 1934 theory of the Wien effect is in all respects a remarkable achievement: a theory of a non-linear non-equilibrium problem of great complexity, with widespread relevance to real problems.
Considering how bound ion pairs (Bjerrum pairs) dissociate into free ions and vice versa: $(+-) \rightleftharpoons (+) + (-)$, Onsager showed that the dissociation constant $K$ changes as an explicit function of electric field $E = |\vec E|$ and temperature $T$. For an electrolyte of charges $\pm q$ and permittivity $\epsilon = \epsilon_0\epsilon_r$, he found the scaling form:
\begin{eqnarray}\label{one}
K(E) / K(0) \equiv F(x) = I_1(\sqrt{8x})/\sqrt{2x} = 1+x+x^2/3+\mathcal{O}\left(x^3\right)
\end{eqnarray}
where 
$x=q^3 E / 8\pi\epsilon (k_\mathrm{B}T)^2$ and $I_1$ is a modified Bessel function. 
The result is universal in that initial concentrations and mobilities disappear from the dimensionless argument $x$, which can be cast as the ratio of two lengths: the Bjerrum length $\ell_T = q^2/8\pi \epsilon k_\mathrm{B} T$ and the field length $\ell_E = k_\mathrm{B}T/qE$ (see Fig. 1).
The leading linear term signals the non-equilibrium nature of the effect, as it would be forbidden by symmetry in equilibrium. It gives a remarkably large response at small fields, making the Wien effect a dominant phenomenon in many experimental situations. 

The increase in $K(E)$ is manifest in the relative change in free ion concentration $n_\mathrm{f}(E) / n_\mathrm{f}(0)$. The strong and weak electrolyte are those limits in which there is an exhausted or an inexhaustible source of charge, respectively. For the strong electrolyte, the relative change $n_\mathrm{f}(E) / n_\mathrm{f}(0)$ is a function not only of $F(\ell_T / \ell_E)$, but also of the initial density. In the weak electrolyte limit, the free charge concentration evolves in a universal manner: $n_\mathrm{f}(E) / n_\mathrm{f}(0) = \sqrt{F(\ell_T / \ell_E) }$, while the bound pair concentration is buffered, i.e.\ replenished rapidly from the reservoir so that $n_\mathrm{b}(E) / n_\mathrm{b}(0) = 1$. 

New charges appear in the form of closely bound pairs 
yielding a double equilibrium between charge source, pairs and free charges: $(0) \rightleftharpoons (+-) \rightleftharpoons (+) + (-)$.
The `vacuum' ($0$) may be a classical electrolyte of molecules which can dissociate into charges, while more general charge vacua 
occur in many chemical and physical processes. Important examples are 
${\rm 2H_2O \rightleftharpoons [H_3O^+ HO^-] \rightleftharpoons H_3O^+ + OH^-}$, 
in the case of both water and ice; 
thermal and optical electron-hole generation in semiconductors~\cite{Pai}; or magnetic monopole generation in spin ice~\cite{CMS,Jaubert}. These effects together with the intrinsic interest in discretizing a continuum model
motivated our study of the Wien effect on a lattice. Notably while ab-initio simulations of real materials have little chance of approaching the Wien effect regime, a simple yet versatile stochastic lattice model is a good starting point for access to this universal non-equilibrium physics.

We present extensive numerical simulations of the Wien effect, which allow us to compute {\it bulk} quantities -- density, conductivity -- at the same time as {\it microscopic} correlations.
The absence of Wien effect simulations in the literature appears in part due to the technical difficulties involved: a combination of low ion densities, strong correlations and long-range interactions pose formidable obstacles. However, by turning to a lattice based simulation, we are able to simulate a regime where the Wien effect is observable -- in fact, strong -- and which is relevant for many problems. 

Monte Carlo simulations were performed on a $1:1$ symmetric electrolyte on a diamond lattice, with periodic boundaries and with the field placed along the $\left[100\right]$ cubic axis (see Methods). The diamond structure was chosen because its four-fold coordination is applicable to many experimental situations, such as proton transport in water and ice, electron transport in germanium and silicon and effective magnetic monopole transport in spin ice materials. 

\begin{figure}[!htbp]
\begin{center}
a) \includegraphics[width=0.47\linewidth]{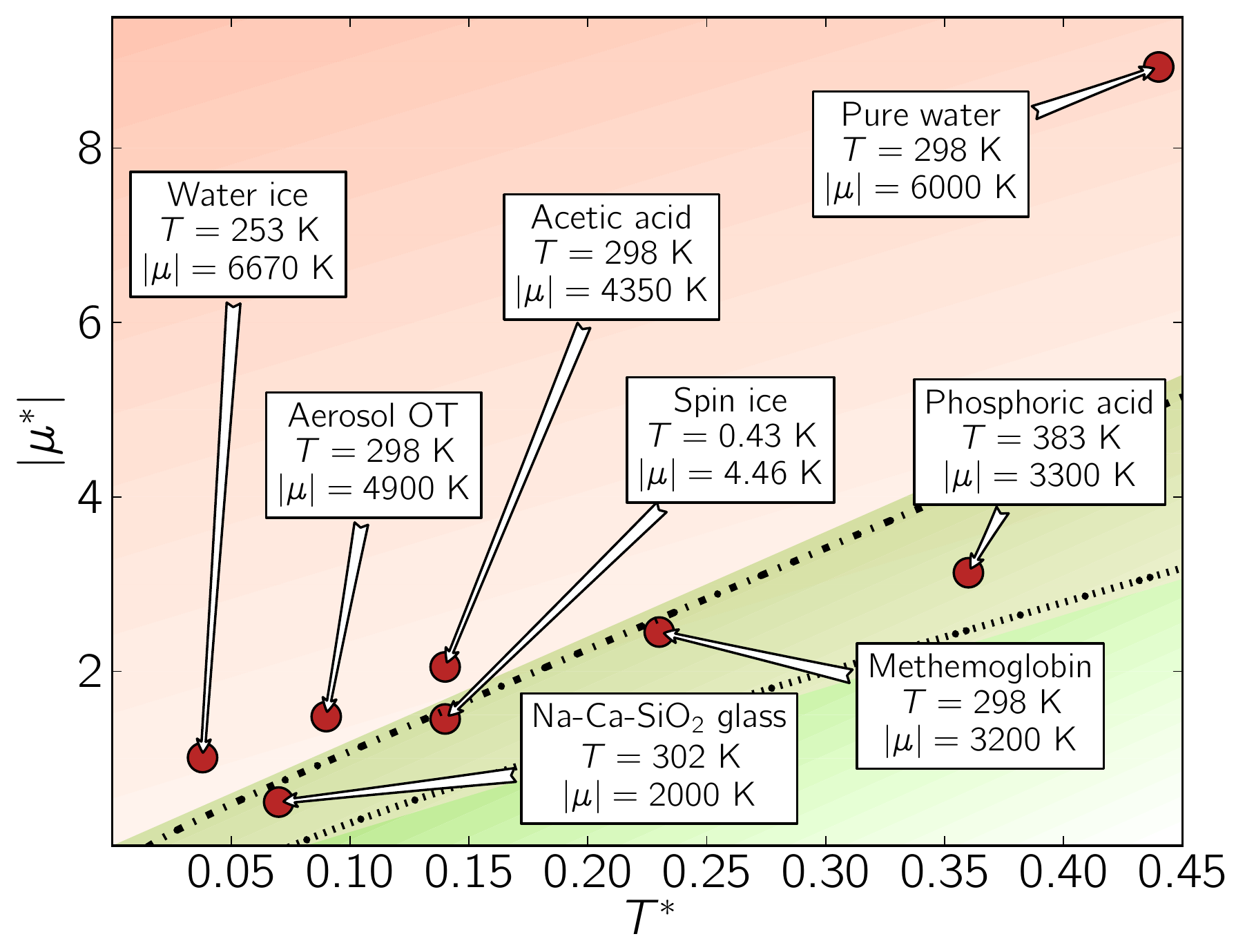}
b) \includegraphics[width=0.46\linewidth]{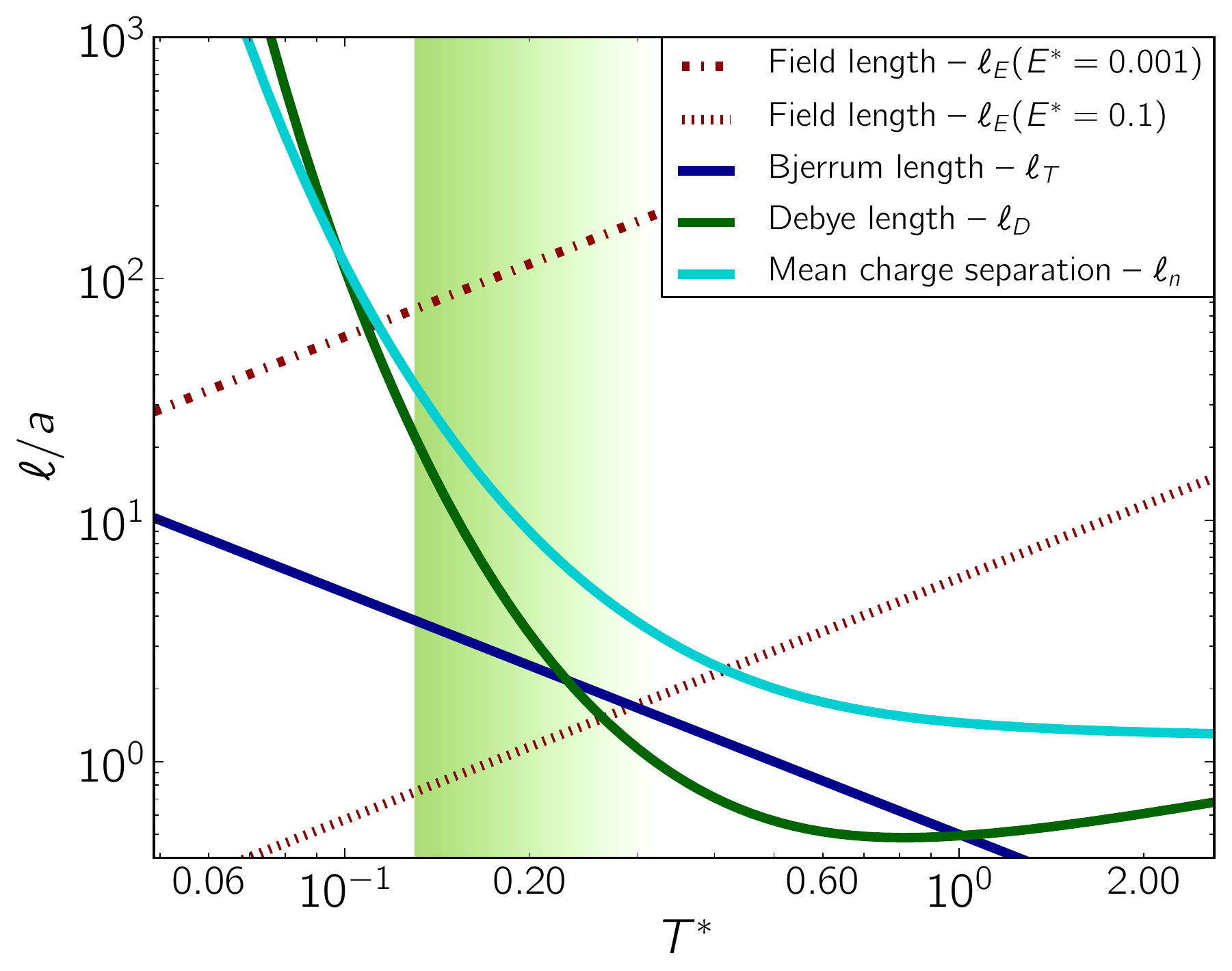}
\caption{
\textbf{a, Reduced variables.} Rough estimates of $|\mu^*|$ and $T^*$ for systems where the second Wien effect has been observed. Water and water ice are strong protonic conductors exhibiting the Wien effect. Molten phosphoric acid -- the strongest known protonic conductor -- was added as it should show interplay of the Wien effect and screening. The Wien effect was first measured in acetic acid. For further details see Supplementary Table 1 and References. The green shaded area is accessible to our simulations. Red shading corresponds to dilute weak electrolyte limit. The dashed line gives the limit of applicability of Onsager's original theory. The dotted line shows validity of theory when extended to include screening.
\textbf{b, Length scales in Coulomb gas for $\mu^* = -1.45$} (relative to diamond lattice site separation). At the Bjerrum length $\ell_T$ (blue line) thermal fluctuations become comparable with the Coulomb binding energy. The Debye length $\ell_\mathrm{D}$ (green line) is the screening length. The mean charge separation $\ell_n$ (cyan line) shows the dilution of the electrolyte. The field length $\ell_E$ give the distance over which the field $E$ exerts work equal to thermal energy. The shaded region shows the extent of our simulations. Extreme dilution of the Coulomb gas prevents simulations at low temperatures, while at high temperatures the electrolyte is dense and highly correlated. The maximum system size simulated was $8 \times 24^3 = 110\,592$ diamond lattice sites corresponding to $L/a = 55.4$.
}
\label{fig:lengthscales}
\end{center}
\end{figure}

The system properties are controlled by two thermodynamic variables: the free charge chemical potential $\mu$ and the temperature $T$. It is appropriate to reduce these by the Coulomb energy at contact $U_0 = q^2 / 4\pi\epsilon a$, to create two dimensionless variables $\mu^* = \mu/U_0$ and $T^* = k_\mathrm{B}T / U_0$. Fig.~1a shows how similar parameters represent a variety of different systems. Fields are likewise given in reduced units, $E^* = U_E / U_0$ where $U_E$ is the work done by field $E$ when a charge hops between neighboring sites. 
Note $E^*$ corresponds to the local field, so that in an experimental situation the depolarizing effects of the contacts on external field would need to be taken into account. However, depolarizing fields from the bulk, through Debye relaxation of bound pairs and the non-Ohmic contribution to the conductivity \cite{BramwellGiblin} should make negligible corrections to $E^*$ at the low densities considered here.
We choose $T^*$ and $\mu^*$ to obtain the dilute weak electrolyte regime $1 \gg n_\mathrm{f} \gg n_\mathrm{b}$. 

Fig.~1b displays the accessible window in which the Wien effect can be analyzed in terms of the Bjerrum and field lengths as well as two other characteristic length scales of the Coulomb gas: the Debye length and the mean charge separation (See Supplementary Discussion 1 for a detailed discussion).

\begin{figure}[!htbp]
\begin{center}
a) \includegraphics[width=0.44\linewidth]{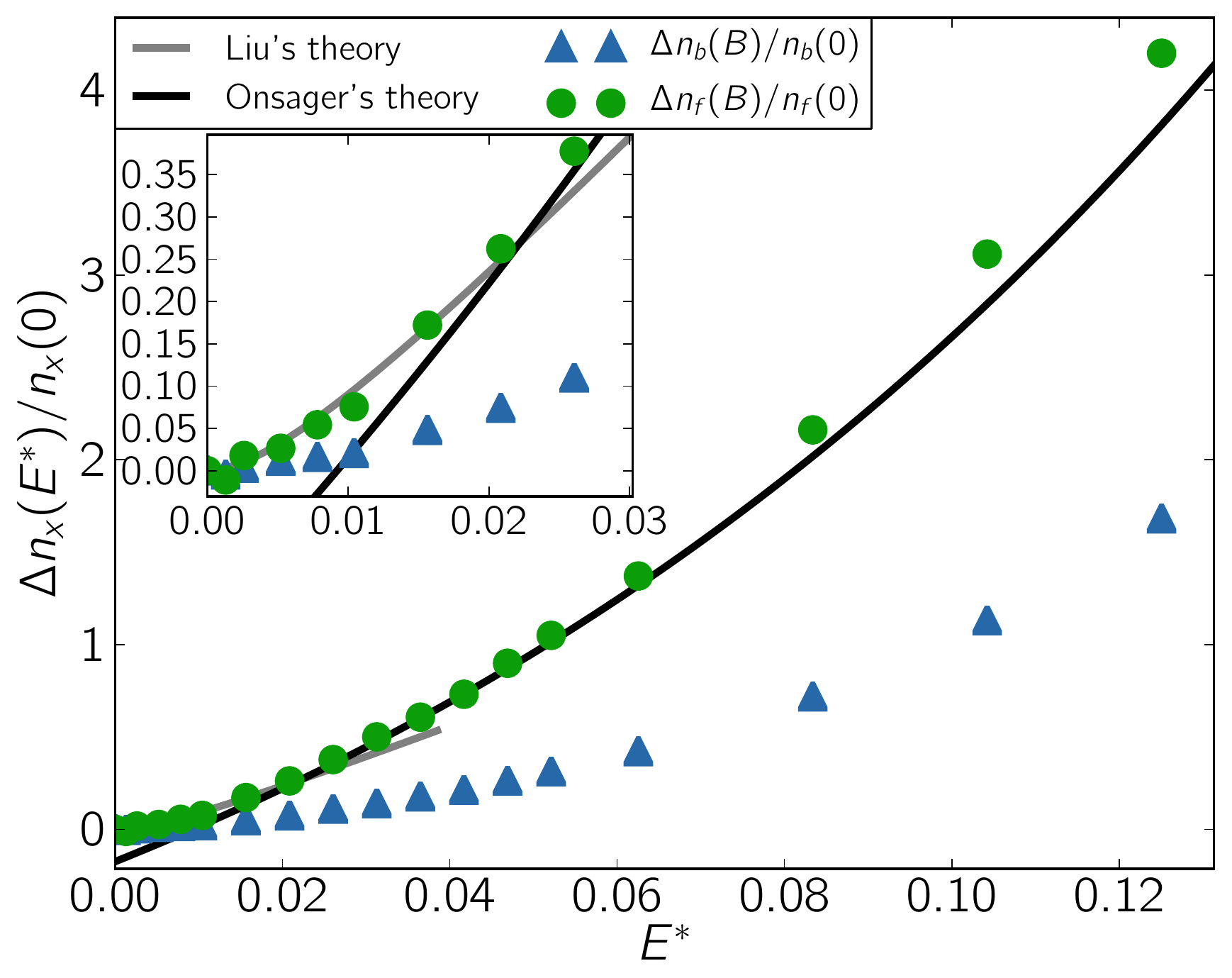}
b) \includegraphics[width=0.48\linewidth]{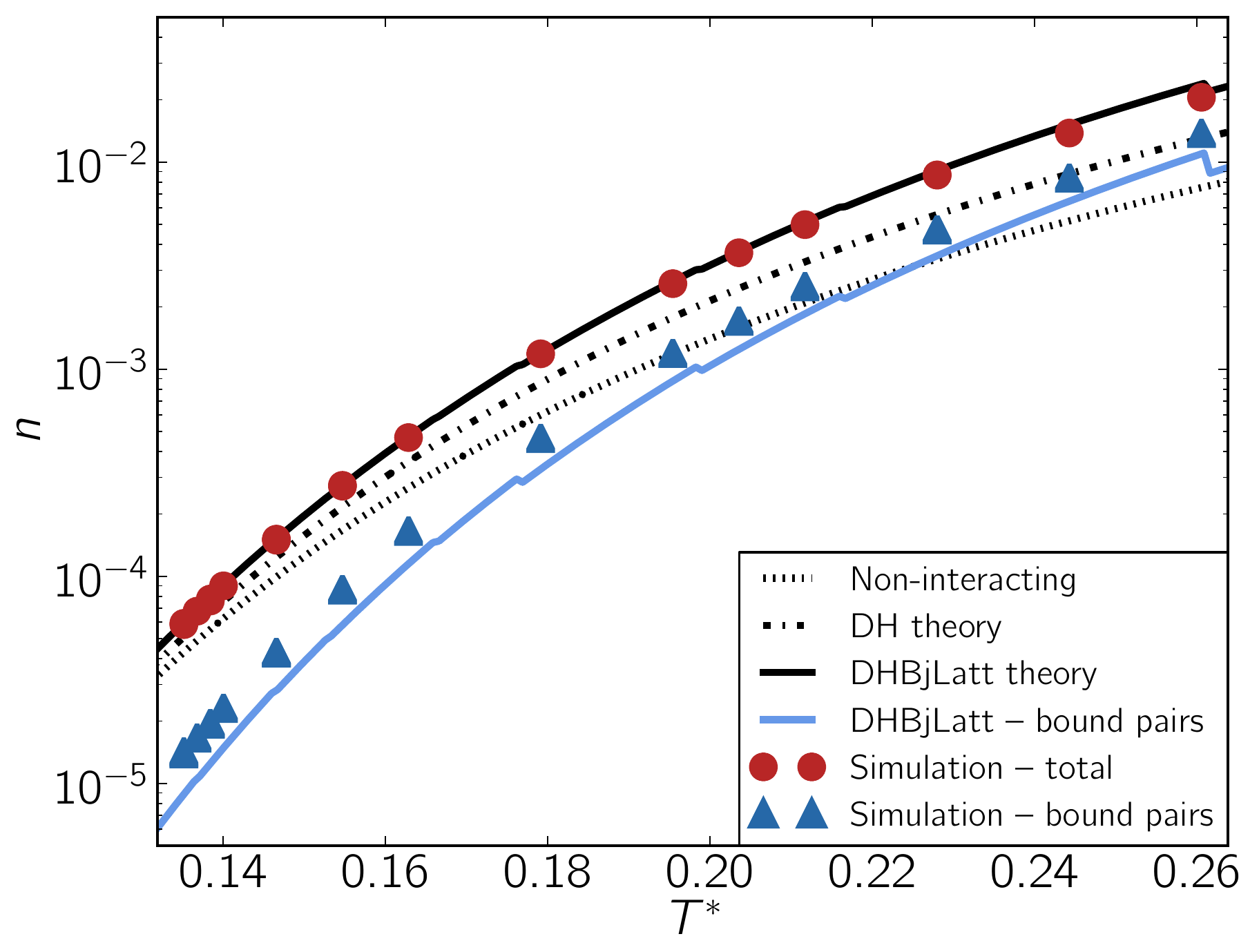}\\
c) \includegraphics[width=0.46\linewidth]{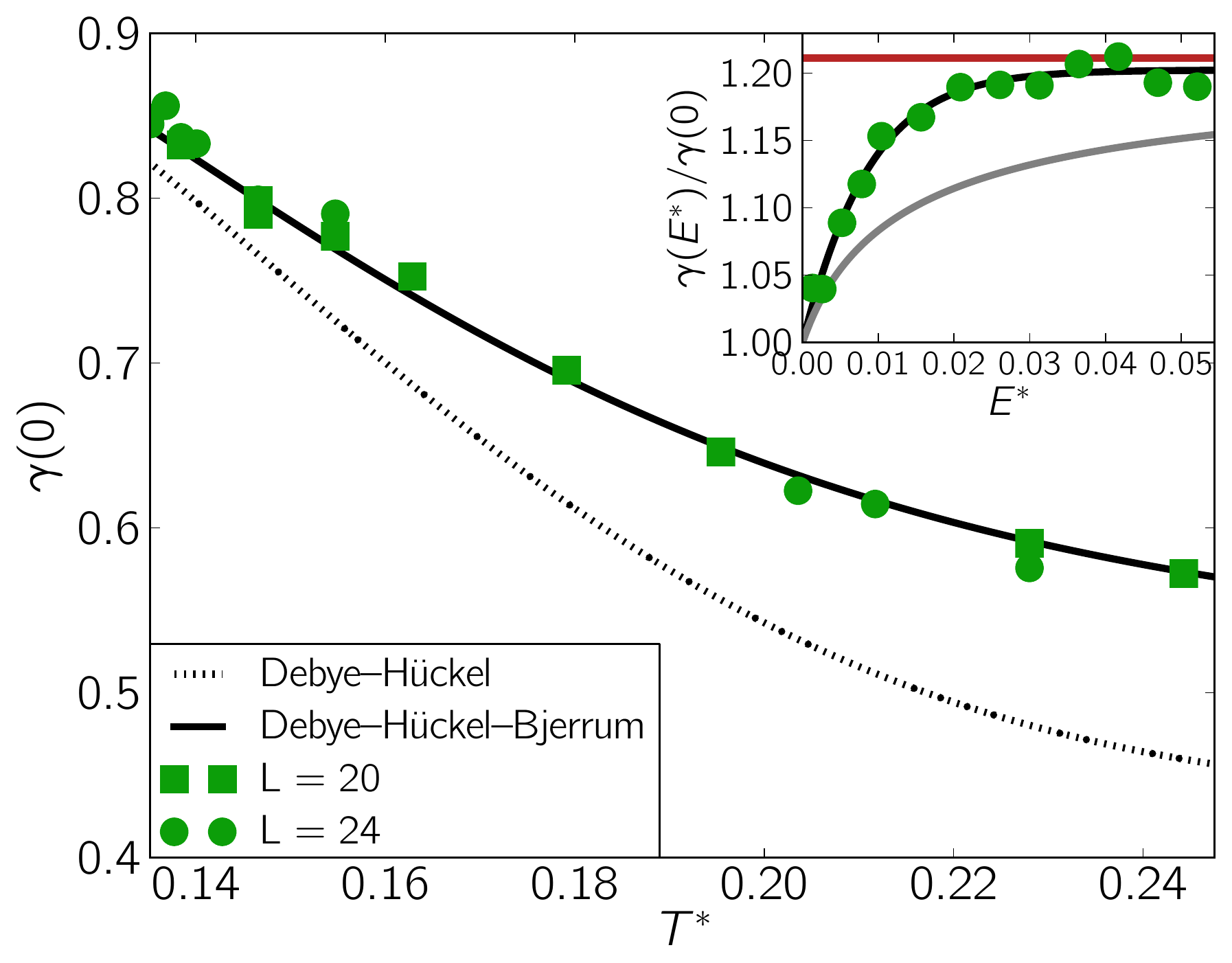}
d) \includegraphics[width=0.46\linewidth]{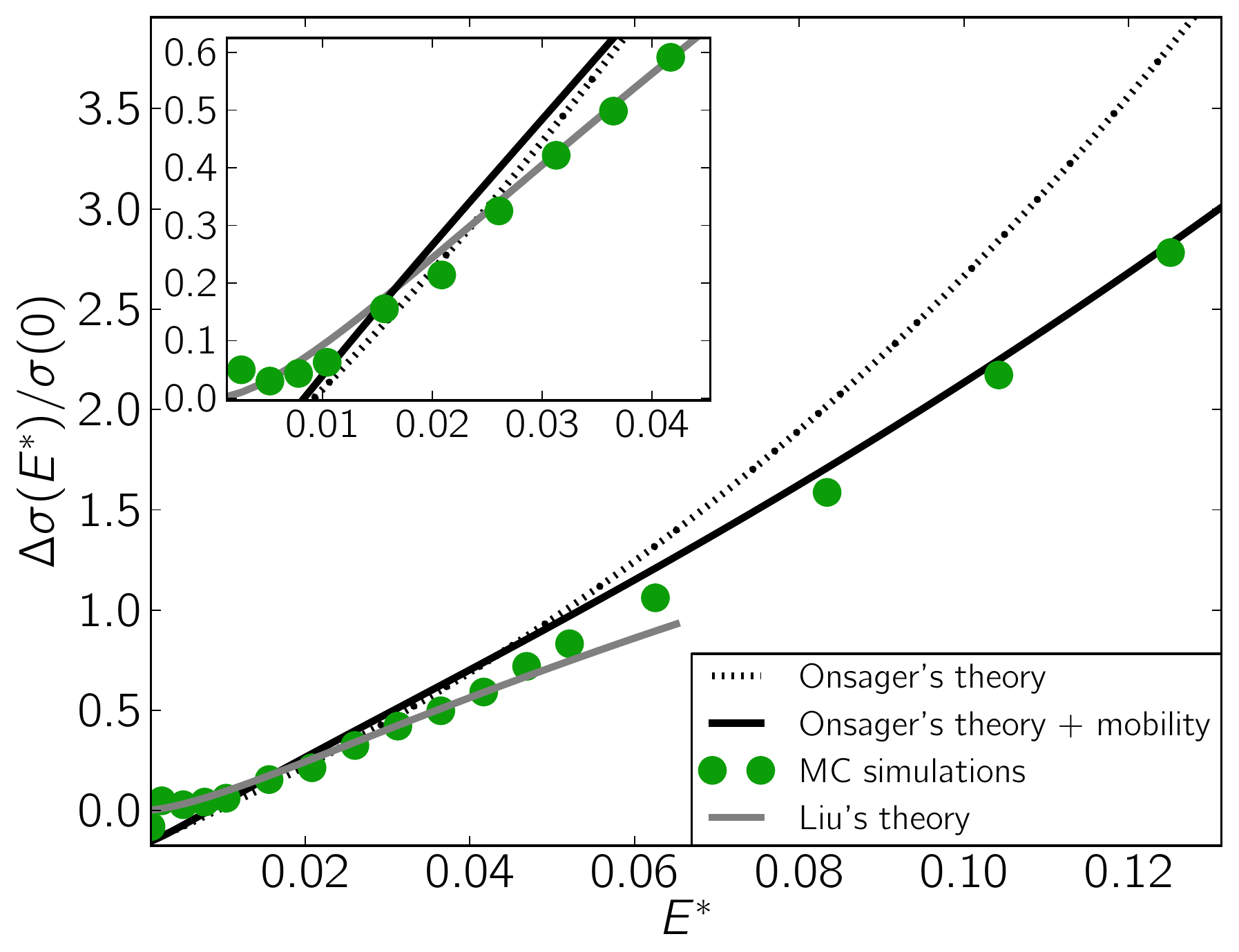}
\end{center}
\caption{
\textbf{Second Wien effect ($|\mu^*| = 1.45$)}. 
\textbf{a,} The charge density increases strongly in field at $T^* = 0.140$. The black curve is Onsager's theory. The linear regime is visible for $E^*$ below 0.04. Onsager's theory extrapolates to a negative intercept on the ordinate, $\gamma(0)-1$. The breakdown of screening occurs when $\ell_E$ exceeds $\ell_\mathrm{D}$, and the intercept on the abscissa corresponds to the field where the two lengths are equal, at low temperature. Liu's theory (grey curve) accounts for screening in the lowest fields. At high fields, finite size effects appear (see Supplementary Discussion 6).
\textbf{b,} The total zero-field density (circles) is well modelled by Bjerrum theory (full black line). Simpler theories, including Debye--H\"uckel (dashed line) and non-interacting theory (dotted line), underestimate the density. The extracted bound pair density (triangles) is slightly underestimated by Bjerrum theory (blue line), but the temperature dependence is correct. Kinks on the lines appear because the lattice association constant is discontinuous where the Bjerrum length crosses subsequent lattice site separations.
\textbf{c,} Zero-field free-charge activity coefficient $\gamma(0)$ extracted from the density increase matches predictions of Bjerrum theory. Comparison with Debye-H\"uckel theory neglecting bound pairs is also shown. Inset: The quality of the fit $\gamma(E^*)/\gamma(0)$ remains good for all temperatures shown, confirming that the Wien effect survives with appropriate corrections up to high temperatures and densities $\sim 10^{-2}$. Extracted $\gamma(E^*)/\gamma(0)$ for $T^*=0.140$ is compared with Liu's function (grey line) and a fitted exponential decay of correlations (black line). Red line denotes the theoretical value of $1/\gamma(0)$.
\textbf{d,} The conductivity increase is similarly significant and also involves field dependence of the ionic mobility. It is reduced in the case also involving field dependence of Metropolis dynamics. The black curve represents Onsager's theory adjusted for both Metropolis dynamics and the influence of the screening cloud on mobility.
}
\label{fig:wien_effect}
\end{figure}

The evolution of $\Delta n_\mathrm{f}/n_\mathrm{f}$ with field at fixed temperature $T^*=0.140$ is shown in Fig.~2a. The first thing to notice is that there is a huge effect! A five-fold increase in the free charge density is induced for a field as small as $E^*=0.125$, even though it provides the smallest energy scale in the problem. 
The second is that there is close agreement between numerics and theory. The data are compared with both Onsager's original theory and the subsequent modification, which takes into account Debye screening effects~\cite{Onsager,Liu}, the latter being characterized by the activity coefficient $\gamma$. Coulomb interactions shift the effective chemical potential by $k_\mathrm{B} T \ln \gamma$. The dissociation constant, $K$ is related to the densities through the concentration quotient, $K_\mathrm{c} = (n_\mathrm{f}/2)^2 / (n_\mathrm{b}/2)=K/\gamma^2$, so that if $n_\mathrm{b}$ is independent of field
\begin{equation}\label{two}
\frac{\Delta n_\mathrm{f}(E)}{n_\mathrm{f}(0)} = \sqrt{\frac{K_\mathrm{c}(E)}{K_\mathrm{c}(0)}} - 1 = \frac{\gamma(0)}{\gamma(E)} \sqrt{F(\ell_T / \ell_E) } - 1.
\end{equation}
In the first case, one assumes that the Debye screening cloud surrounding a charge is destroyed by the applied field, so that $\gamma(E)=1$. This describes all data except for the lowest fields ($E^* = 0.02$). For smaller fields Liu \cite{Liu} proposed a logarithmic field dependence for $\gamma(E)$ (see methods).  From Fig. 2a (inset) one can see that we find close agreement with Liu's theory in the low field region. At higher fields the theory underestimates the observed Wien effect by around $5\%$. This difference can be traced to finite size effects in the simulation (see Supplementary Discussion 6). We note that in making these comparisons, there are no fitting parameters, as the activity coefficient is a quantity calculated independently (see Methods). 

Also shown in Fig. 2a, is the relative increase in bound pair density as a function of field, $\Delta n_\mathrm{b}/n_\mathrm{b}$.  Bound charges are defined as being separated by less than the Bjerrum length, but we find that the functional form is  independent of the details of this definition~\cite{McIlroy}; $n_\mathrm{b}$ does indeed remain approximately constant up to $E^*=0.04$, hence confirming the buffering hypothesis. Above this threshold there is an increase in bound pair concentration, but at a reduced rate compared with $\Delta n_\mathrm{f}/n_\mathrm{f}$. The field therefore generates a considerable increase in the total charge concentration.

The activity coefficient used in the comparison was calculated using Debye--H\"uckel--Bjerrum theory, which regards free charges and bound pairs as distinct chemical species. The quality of this approximate theory for the values of $T^*$ and $\mu^*$  chosen, is illustrated in Fig 2b, where we show the evolution of the total density and of $n_\mathrm{b}$ as a function of $T^*$ for zero external field. The density is predicted accurately and $n_\mathrm{b}$ to a reasonable approximation. In Fig 2c we compare the calculated and measured values $\gamma(0)$ for the free particles, which in the dilute regime can be expressed as the ratio of ideal gas density and actual density: $\gamma(0) = 2\exp(\mu^*/T^*) / n_\mathrm{f} < 1$. Numerics matches theory for the Wien effect to below $\gamma(0) = 0.6$, well beyond the formal limit of validity $\gamma(0) \rightarrow 1$. In Fig. 2b (inset) we show the evolution of $\gamma(E)$ for $T^*=0.140$, this time extracted from simulation using equation (2). The crossover to an asymptotic regime with $\gamma(E)\sim 1$ occurs for $E^*\sim 0.02$, as can be expected from Fig. 2a. While Liu's expression captures the initial increase of $\gamma(E)$, it turns out that an exponential decay of screening captures the crossover better\cite{Liu}.

The Wien effect was originally observed in measurements of conductivity, $\sigma(E)$, which exhibits large non-Ohmic corrections to the zero field value, $\sigma(0)$~\cite{Onsager}. 
In many of the systems listed above, $\sigma=q^2\omega n_\mathrm{f}$, is of great practical importance. It can be used as a diagnostic for the Wien effect if the mobility, $\omega$, is field independent, or if its field dependence is known.
In Figure 2d we show the relative increase in conductivity from numerical simulation at $T^* = 0.140$. Again we remark that the increase is large; the conductivity triples over the field range studied. However, the increase falls systematically below that predicted from equation (\ref{two}). 

\begin{figure}[!htbp]
\begin{center}
\includegraphics[width=0.5\linewidth]{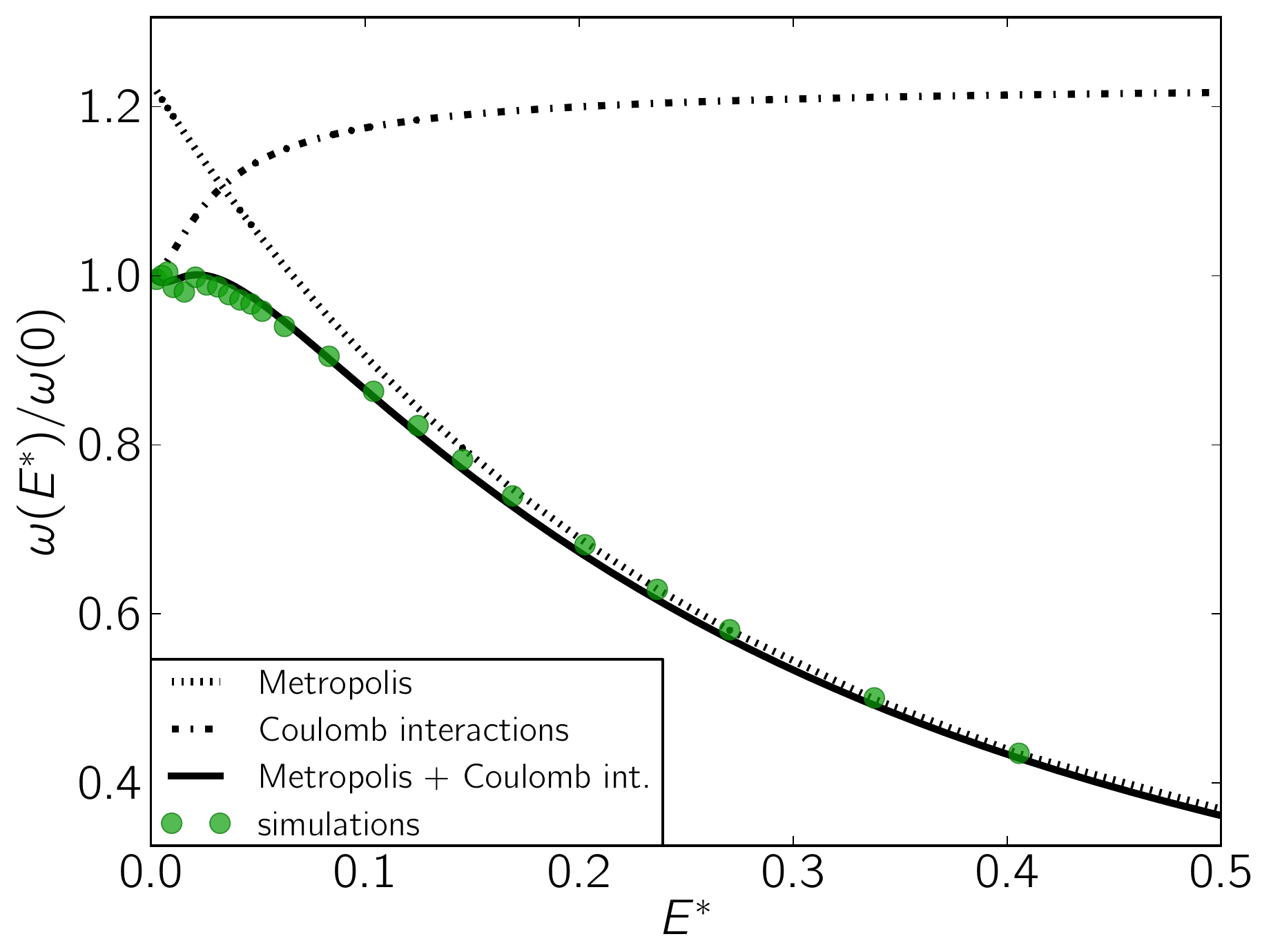}
\end{center}
\caption{
\textbf{Relative mobility change with field ($\mu^* = -1.45$, $T^* = 0.155$)}. The high field mobility is dominated by a reduction characteristic of Metropolis dynamics employed in our simulations, while at low fields the interactions influence the mobility. Mobility at lowest fields is noisy as the drift speed becomes very small.}
\label{fig:mobility}
\end{figure}

The shortfall arises because the mobility is a decreasing function of field, as shown in figure \ref{fig:mobility} for Metropolis dynamics. For non-interacting particles, Metropolis dynamics yields: $\omega_0(E^*) / \omega_0(0) = (1 - \exp(- E^* / T^*)) / (E^* / T^*)$.
For strong fields, the simulated mobility clearly approaches this asymptotically, while for small fields it is Coulomb interactions which lead to a reduced value. A crossover function can be constructed by first estimating the relative zero-field reduction, $h(\ell_T,\ell_\mathrm{D})$, from Fuoss-Onsager theory for the conductivity\cite{FuossOnsager} and then estimating the suppression of this reduction by the field, $g(\ell_\mathrm{D}/\ell_E)$ from Wilson's theory for the first Wien effect.
(details in Supplementary Discussion 5):
\begin{equation}
\frac{\omega(E^*)}{\omega(0)} = \frac{1 - h\left(\ell_T, \ell_\mathrm{D}\right) g\left(\ell_\mathrm{D} / \ell_E \right)}{1 - h\left(\ell_T, \ell_\mathrm{D}\right)} \frac{1 - \exp(- E^* / T^*)}{E^* / T^*}.
\end{equation}
The data in Fig. \ref{fig:mobility} accurately follows this expression. Supplementing the Wien effect theory (\ref{two}) with an analysis of the field dependent mobility restores a good agreement of theory and simulations in Fig. 2d.

Over and above the numerical verification of the existence of the Wien effect we see here that simulation can provide deep insight into the role of the microscopic dynamics for this non-Ohmic phenomenon, providing specific information on the field dependent mobility both in model and in experimental systems. Simulations such as ours can thus be used for the development of models for microscopic dynamics. 

\begin{figure}[!htbp]
\begin{center}
a) \includegraphics[width=0.46\linewidth]{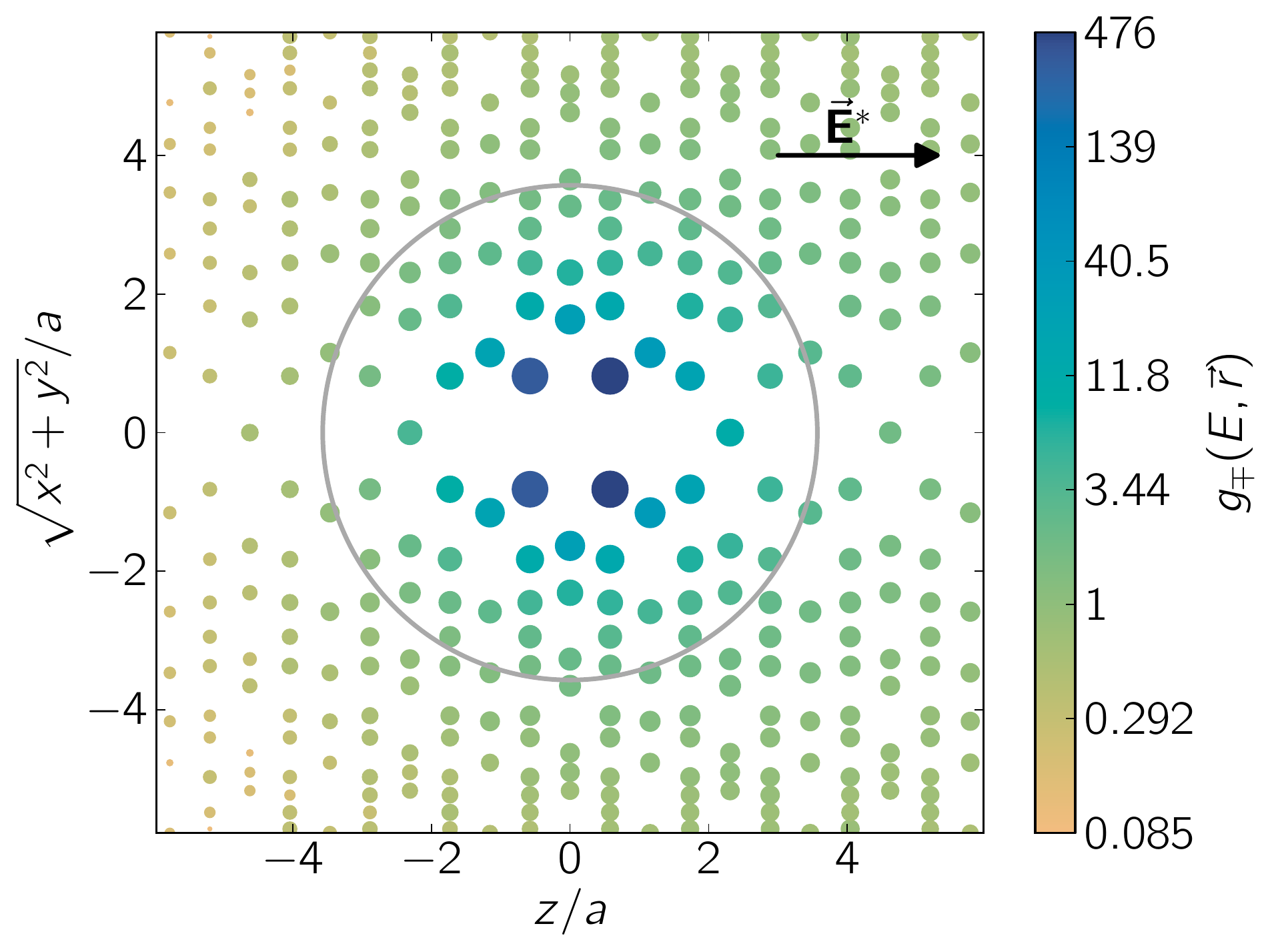}
b) \includegraphics[width=0.47\linewidth]{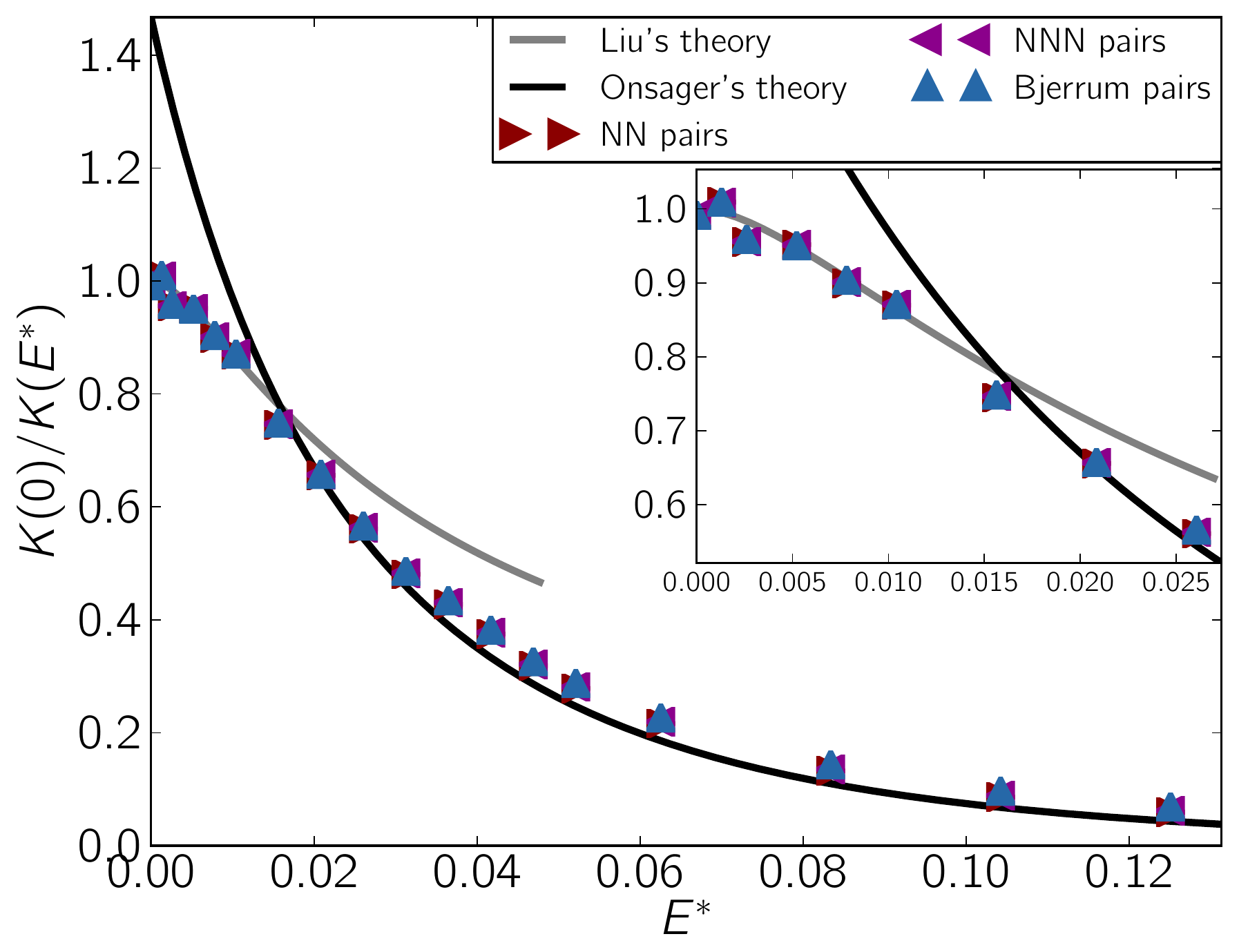}
\end{center}
\caption{
\textbf{Charge correlations ($\mu^* = -1.45$, $T^* = 0.140$)}. 
\textbf{a,} Pair correlation function ($g_{\pm}$) for $E^* = 0.021$ along the $\hat{z}$ axis; averaged over the azimuthal angle and mirrored for clarity. The statistical weight is strongly peaked at nearest-neighbour sites, which are most affected by the lattice structure, while asymmetry develops in the field. The black circle marks the Bjerrum volume.
\textbf{b,} The decrease in the ion association $1/K_\mathrm{c} = \sum_{|\vec r| < \ell_T} g_{\pm}$. The Wien effect leads to reduction of correlations initially linear with field as the charges escape bound pairs. In the low field region the quadratic field dependence is recovered due to screening. Different symbols show that restricting the bound pairs from the Bjerrum volume to nearest neighbours or up to second-nearest neighbours yields similar behaviour because the nearest-neighbour contribution dominates the correlations. The Onsager function is common to all definitions of bound pairs, while $\gamma(0)$ differs slightly for different definitions of free charges, which can be best observed at high fields. The zero-field value of the association constant differs from the continuum value found by Bjerrum, but their relative change is universal.
}
\label{fig:charge_correlation}
\end{figure}

Complementing macroscopic quantities, the microscopic processes at the heart of the Wien effect are also accessible to our simulations. Particularly important is $g_{\pm}(\vec r)=(\left<n^+(\vec r)n^-(0)\right>-\left<n^+(\vec r)\right>\left<n^-(0)\right>)/\left<n^+\right>\left<n^-\right>$, the correlation function between a negative charge at the origin and a positive charge at position $\vec r$, where $\left<\cdot\right>$ denotes thermal average. The pair correlation function is shown in Fig. 4a, where in finite field an asymmetry develops along the field axis. The amplitude of the correlations at small $r$ decreases strongly with increasing field.

The correlation function divides into symmetric and anti-symmetric parts,  $g_{\pm}^\mathrm{S/A}(\vec r)$, with respect to inversion $\vec r \rightarrow -\vec r$ or $\vec E \rightarrow -\vec E$. 
The sum of $g_{\pm}^\mathrm{S}(\vec r)$ over $|\vec r| < \ell_T$ gives access to the observables of the Wien effect: $\sum_{|\vec r| < \ell_T} g_{\pm}^\mathrm{S}(\vec r) = 1/K_\mathrm{c}$. This is confirmed in Fig 4b where we compare the evolution of $K(0)/K(E^*)$ evaluated directly from $g_{\pm}^\mathrm{S}(\vec r)$, with the theoretical expressions. An agreement is found both in the low field, screened region \cite{Liu}, and the higher field region \cite{Onsager}, thereby linking microscopic correlations with bulk (experimental) observables.

Having established a simple stochastic model that exhibits the Wien effect in both the low and high density regimes future work should identify the extent to which this universal mesoscale physics is supported by the detailed microscopics of specific material systems. Problems of interest concern, for example,
the competition of Brownian diffusion with variable range hopping and the various injection processes in Poole-Frenkel conduction in semiconductors~\cite{Pai}. Here the possibility of a Wien effect for fractionalized carriers on `frustrated' lattices~\cite{Fulde} can also be tested. A second example concerns the generation of photocurrents in solar cell devices, varying the role of an exciton as a short lived transient or as as an effective ion pair~\cite{Braun}. Other examples include ionic conduction in oxide glasses~\cite{Tomoz}, where there is experimental evidence of the Wien effect, but where non-rigid networks, disorder and rough energy landscapes are complicating factors; protonic conduction in hydrogen-bonded networks like water ice or phosphoric acid~\cite{Vilc}, where correlated Grotthus-type conduction competes with other processes; or magnetic monopole conduction in spin ice~\cite{RyzhkinSpinIce,BramwellGiblin} or artificial spin ice~\cite{WangSchiffer,Branford}, where similar constraints apply and where one also has to contend with the effects of large amplitude internal fields \cite{Dunsiger}.

In all these specific cases, it should be possible to justify the stepwise addition of system-specific features to our basic model, so that future simulations may be adapted to incorporate a diverse range of system details, such as competing dynamical processes, disorder, finite size and boundary conditions~\cite{Canadians, Ryzhkin}. As a first example we lay down in the SI, steps for the development of a better understanding of high field conduction in water ice, directly linking our model to a more involved microscopic description. In this way, the advantages of our approach, including direct access to correlation functions and the processes that drive the chemical equilibria, may be exploited to understand and reveal new aspects of the Wien effect in practical material systems. That there is a need for such a programme has been recognised in the literature~\cite{DuhkinParlia}, and it is noteworthy that an issue clarified in this work -- the correction for field mobility -- has previously been flagged as a problem~\cite{Ingram,Tomoz}.

More generally, the advantages offered by finite lattice simulations could be exploited to assess the effect of a vanishing number of ion pairs, as might be relevant to liquids in the highly confining environment of biophysical systems, or else to explore the influence of the Wien effect on films confined between membranes, or the electrical double layer~\cite{Bruesch}, where large gradients of electrical potential can be created. In this context dynamical studies of the Wien effect would also be relevant~\cite{Persoons}. Thus our method may in the future be specialized to treat outstanding problems of non-linear conduction across a wide variety of scientific fields, from electrochemistry and biochemistry to solid state physics and electrical engineering.

\newpage
{\bf \large Methods}

\textbf{The zero-field concentration} 
can be computed iteratively from the following set of equations:
\begin{eqnarray}
n &=&  \frac{2\exp\left(\mu^* / T^*\right)}{\alpha\gamma + 2\exp\left(\mu^* / T^*\right)} \\
\gamma &=&  \exp\left(- \frac{\ell_T}{\ell_T + \ell_\mathrm{D} / \sqrt{\alpha}}\right) \\
\alpha &=&  \frac{2}{1 + \sqrt{1 + 2 K_A \gamma^2 n}} \\
\ell_\mathrm{D} &=&  \sqrt{\frac{1}{8 \pi \ell_T n}} \,,
\end{eqnarray}
where the first equation gives the density of a lattice gas with chemical potential $\mu^*$, degree of dissociation by $\alpha = n_\mathrm{f} / (n_\mathrm{f} + n_\mathrm{b})$, and free charge correlations described by $\gamma$. The activity coefficient $\gamma$ follows the Bjerrum mean field theory, while $\alpha$ results from the chemical equilibrium between free and bound charges given by association constant $K_A$. The last equation is the definition of the Debye length.
The Bjerrum association constant $K_A$ is obtained by lattice summation:
\begin{equation}
K_A = \sum_{\{\vec r : r \leq \ell_T \}}\exp\left(- \frac{2\ell_T}{r}\right)
\end{equation}
Using the continuous approximation to $K_A$ overestimates the concentration at low temperatures and underestimates it at high temperatures. Concentrations are related to the volume per site and correspond to molar fractions of individual species. Association and dissociation constants are related by $K_A = 1 / K$.

\textbf{Liu's formula:} In the case where screening dominates the Wien effect ($l_E \gg l_D$), Liu and Onsager solved perturbatively the two body dynamics in the Debye potential with the first order result:
\begin{equation}\label{Liueq}
\frac{K_\mathrm{c}(E)}{K_\mathrm{c}(0)} = 1 + \frac{\ell_T / \ell_E + (2\ell_T / \ell_\mathrm{D}) f(\ell_\mathrm{D} / \ell_E)}{1 + 2\ell_T / \ell_\mathrm{D}} + \mathcal{O}\left(\left(\frac{\ell_T}{\ell_E}\right)^2\right), \quad f(x) = \frac{\ln(1 + x)}{x}\,.
\end{equation}
Note that this formula only captures the linear term $\ell_T / \ell_E$ of the Wien effect and uses the following approximation of the activity coefficient $\gamma = \exp(- \ell_T / (\ell_T + \ell_\mathrm{D} / \sqrt{\alpha})) \simeq 1 - \ell_T / \ell_\mathrm{D}$. To this degree of approximation the activity coefficient in field is\cite{Liu}: $\gamma(E) = \gamma(0)^{f(\ell_\mathrm{D} / \ell_E)}$.

\textbf{Monte Carlo:} Simulated systems were blocks of diamond lattice with $8\times L^3$ sites, with $L$ up to 24 and with periodic boundary conditions. Each site can be empty or occupied by a positive or a negative unit charge. Updates proceed by selecting a bond between two sites chosen at random and creating or annihilating a pair of charges (at fixed chemical potential $\mu^*$), or by moving a charge along the bond. Acceptance probabilities are given by the Metropolis algorithm. The Coulomb interaction between charges is treated by Ewald summation with tinfoil boundary condition. Sample systems are first equilibrated in zero-field, the [001] field is set to a given value. After the transient effects vanishes, which we verify using the density auto-correlation function, the in-field values are extracted. Free charges are separated from bound pairs using $K_\mathrm{c} = n_\mathrm{f}^2 / (2 (n - n_\mathrm{f}))$, where $K_\mathrm{c}$ is extracted from the pair correlation function.  

In our simulations the constant field generates a direct current flowing through the periodic boundaries~\cite{Jaubert}, with energy dissipation of $qLE$ per charge passing through the periodic cell. The fundamental problems of applying an electric field on a torus are bypassed here as the stochastic dynamics of the Metropolis algorithm are intrinsically non-ballistic, which ensures a finite value of $\sigma$. The current is computed as a bulk quantity, therefore any process moving a positive charge in the field direction gives a positive current and vice versa.

{\bf \large Acknowledgements}

We thank L. Jaubert for generously sharing and discussing his numerical code with us. PCWH thanks L. Bocquet for useful discussions. RM thanks C. Castelnovo and S. Sondhi for many discussions and related collaborations. STB thanks S. R. Giblin for related collaborations.

{\bf \large Author Contributions}

VK conducted the simulations. All four authors contributed equally to the formulation and development of the project, as well as to the text of the paper.

\newpage
{\bf \Large Supplementary Information} 

\vspace{0.5cm} 
\noindent
{\bf 1. Supplementary Discussion: Characteristic length scales} 

There are multiple length scales relevant to the physics of an electrolyte in addition to the Bjerrum length~\cite{Bjerrum}, $\ell_T = q^2 / 8\pi\epsilon k_\mathrm{B}T$ and the field length $\ell_E = k_\mathrm{B}T / qE$: the Debye length $\ell_\mathrm{D} = \sqrt{\epsilon k_\mathrm{B} T V_0 / q^2 n}$, above which the Coulomb interaction is screened through the formation of a correlated charge cloud ($V_0$ is the volume per lattice site); the mean charge separation, $\ell_n = (V_0 / n)^{1/3}$; a minimum length $a$ for the separation of two charges, as well as the macroscopic scale $L$ of our numerical simulation box.
The main result of Onsager's theory, equation (1), is valid in the limit that $\ell_\mathrm{D}$ and $L$ exceed $\ell_T$ and $\ell_E$, which in turn must exceed the microscopic length $a$. The small field limit exposing the linear field dependence requires the rigorous separation of scales:
$L\gg\ell_\mathrm{D} \gg\ell_n\gg\ell_E \gg \ell_T\gg a$, which can only be approached for temperatures well below the temperature scale set by the chemical potential, $\mu$. In Fig. 1b we show the evolution of these lengths with temperature for our lattice electrolyte. The window of application available to our simulations, in which the Wien effect can be measured, is bounded from above by the crossover, $\ell_\mathrm{D} < \ell_T$ and from below by the finite size limit, $L < \ell_n$, below which we have, on average fewer than one pair of particles in our simulation box.

\vspace{0.5cm}
\noindent
{\bf 2. Supplementary Discussion: Reduced units}

In the main text, reduced units for corresponding states of electrolytes were introduced: $\mu^* = \mu/U_0$ and $T^* = k_\mathrm{B}T/U_0$, $E^*=U_E/U_0$. In Supplementary Table 1 we list appropriate parameters for ${\rm Dy_2Ti_2O_7}$ spin ice at 0.5 K, for water ice at 253 K, for pure liquid water at 298 K and several other materials. We added phosphoric acid, since it is the strongest known protonic conductor and should exhibit a screened version of Wien effect (with defect concentrations $\sim0.07$).
Note that the relative permittivity for water ice is not the bulk value but instead is the high frequency value, as advocated by Onsager and Dupois \cite{OnsagerDupuis}, from which other parameters are taken. The low frequency value was assumed for liquid water and phosphoric acid. 
For the spin ice/lattice electrolyte parameters, the corresponding charge concentration of order $n \sim 10^{-5}$ is possible to handle numerically, while being sufficiently small to represent the low density limit in general. Likewise,  Debye--H\"uckel and Bjerrum corrections for this parameter set are small but non-negligible. Hence it is a suitable  system on which to test the expectations outlined above. 

\begin{center}
\vspace{1cm}
\noindent
\begin{tabular}{|c||c|c|c||c|c||c|c|}
\hline
\multicolumn{8}{|c|}{\bf Supplementary Table 1: Parameters for the double equilibria}\\
\hline
\hline
{\bf System} & a [\AA] & $q/e$ & $\epsilon_r$ & $T$ [K] & $|\mu|$ [K] & $T^*$ &$\mu^*$\\
\hline 
Spin ice ${\rm Dy_2Ti_2O_7}$ & 4.33 & 1/112 & 1 & 0.5 & 4.46 & 0.163 & 1.453\\
\hline 
Water ice & 2.75 & 0.58 & 3.1 & 253 & 6670 & 0.038 & 1.01\\
\hline
Pure water & 3.01 & 1 & 80.1 & 298 & 6000 & 0.44 & 8.93\\
\hline
Acetic acid~\cite{Gledhill} (aq) (1 mmol/dm$^3$) & 0.98 & 1 & 80.1 & 298 & 4350 & 0.140 & 2.06\\
\hline
AOT\cite{RanS} in cyclohexane (0.1 mmol/dm$^3$) & 25 & 1 & 2.02 & 298 & 4900 & 0.090 & 1.48\\
\hline
Anhydrous orthophosphoric acid~\cite{VilcS} & 2.6 & 1 & 61 & 383 & 3300 & 0.36 & 3.13\\
\hline
Na-Ca-SiO$_2$ glass~\cite{IngramS} & 5 & 1 & 8 & 302 & 2000 & 0.07 & 0.5\\
\hline
Methemoglobin~\cite{Braeunig} (83 $\mu$mol/dm$^3$) & 1.6 & 1 & 80.1 & 298 & 3200 & 0.23 & 2.74\\
\hline
\end{tabular}
\end{center}
\vspace{0.5cm}

In this study, we simulated a system corresponding to ${\rm Dy_2Ti_2O_7}$ spin ice parameters between 0.41 K and 0.8 K. We also simulated a lattice electrolyte with parameters \{$a = 5$ \AA, $q/e = 1$, $\epsilon_r=20$\}, chemical potentials ranging from 2500 K to 2900 K ($T^* =$ 0.162--0.221), and at temperatures between 270 K and 370 K ($|\mu^*| =$ 1.496--1.735). All systems confirm the conclusions presented in the main text.

\vspace{1cm}
\noindent
{\bf 3. Supplementary Discussion:  Non-equilibrium quantities}

Conceptually, it is important to distinguish between thermodynamic quantities in equilibrium and their extension to the field-driven case. Practically, the only observable out of equilibrium is the concentration quotient $K_\mathrm{c}(E) = n_\mathrm{f}(E)^2 / (2 n_\mathrm{b}(E))$. If we relate it to the dissociation constant and the activity coefficient in the same way as in equilibrium, $K_\mathrm{c}(E) = K(E)/\gamma^2(E)$, we face the problem of separating the increase in free charge density -- because the field increases the dissociation -- from the decrease in free charge density -- due to the reduced screening.

The conductivity measurements in strong electrolytes show that the ionic atmosphere cannot establish itself in strong external fields (first Wien effect). It is safe to assume $\gamma(E) \rightarrow 1$ in high fields. Practically, strong fields in this context corresponds to the length scale ordering $\ell_E < \ell_\mathrm{D}$. In this limit Onsager's prediction matches the simulations and experiments well. Onsager's expression for $K(E)$ is justified and $\gamma(E)$ goes from $\gamma(0)$, as predicted by Bjerrum's theory, to high field limit of unity. 

An early discussion of non-equilibrium quantities in the context of the second Wien effect can be found in Patterson and Freitag~\cite{Patterson} where the ``un-screening coefficient'' was proposed as a designation for the non-equilibrium activity coefficient.

\vspace{0.5cm}
\noindent
{\bf 4. Supplementary Discussion: Buffering}

Two types of double equilibrium can be distinguished by the source of charges: dissociating molecules or creation from the vacuum. The creation from vacuum can also be modelled as a ``reaction'' of two empty sites in case of ice and spin ice. In this section, we present a simple calculation to explicitly show how the double equilibrium shifts due to the second Wien effect. The result for dissociating molecules is stated in Onsager's paper \cite{OnsagerS} without calculation. Charge generation exhibits similar properties. Both cases show buffering of the bound charge concentration in the dilute weak electrolyte limit $n_e \gg n_\mathrm{f} \gg n_\mathrm{b}$.

In the following, $n_m$, $n_e$, $n_\mathrm{f}$, and $n_\mathrm{b}$ are the densities of molecules, empty sites, free ions, and bound ions, respectively. Densities after the equilibrium shifts are denoted with a prime, e.g. $n_\mathrm{f}'$.

The double equilibrium $(0) \rightleftharpoons (+-) \rightleftharpoons (+) + (-)$ where charges are created from molecules is described by:
\begin{eqnarray}
n_\mathrm{b}/2 &=& K_m n_m \\
(n_\mathrm{f}/2)^2 &=&  K_\mathrm{c} n_\mathrm{b}/2 \\
2 n_m + n_\mathrm{b} + n_\mathrm{f} &=&  1 \,,
\end{eqnarray}
where we have normalized the total concentration to unity (mole fraction). Substituting the first two equations into the last gives:
\begin{equation}
2 (n_\mathrm{f}/2)^2 (1 + 1 / K_m) / K_\mathrm{c} + n_\mathrm{f} - 1 = 0 \,,
\end{equation}
which is equivalent to the equation for free charge density in case of a single equilibrium $(+-) \rightleftharpoons (+) + (-)$ with dissociation concentration quotient $K_d = K_\mathrm{c} / (1 + 1 / K_m)$. The Wien effect gives for the dissociation constant in field $K_d(E) = K_d \gamma(0)^2 F(\ell_T / \ell_E)$:
\begin{equation}
\gamma(0)^2 F(\ell_T / \ell_E) n_\mathrm{f}^{\prime 2} / 2 + n'_f - 1 = 0 \,.
\end{equation}
Expressing $n'_f$ in terms of the equilibrium concentration gives:
\begin{equation}
\frac{n'_f}{n_\mathrm{f}} = \frac{\gamma(0)^2F(\ell_T / \ell_E) n_\mathrm{f}}{2n_\mathrm{b}} \left( \sqrt{1+\frac{4(1-n_\mathrm{f})}{\gamma(0)^2 F(\ell_T / \ell_E)n_\mathrm{f}^2}} - 1 \right)
\end{equation}
If $n_\mathrm{f} \ll 1$ , we can expand into series:
\begin{equation}
\frac{n'_f}{n_\mathrm{f}} = \gamma\sqrt{F(\ell_T / \ell_E) } - \gamma(0)^2F(\ell_T / \ell_E) + \mathcal{O}\left(n_\mathrm{f}^2\right) = \gamma\sqrt{F(\ell_T / \ell_E) } + \mathcal{O}\left(n_\mathrm{f}\right) \,.
\end{equation}
The total concentration of bound pairs and molecules $(1 - n_\mathrm{f}) \simeq 1$ stays essentially unchanged as well as the equilibrium between them. The bound pair concentration is constant to first order:
\begin{equation}
\frac{n'_b}{n_\mathrm{b}} = 1 + \mathcal{O}\left(n_\mathrm{f}\right) \,.
\end{equation}
The bound pairs in this case can be considered as precursors of the fully associated molecules. The Wien effect is a non-equilibrium process where the actual rates whose ratio gives the dissociation quotient matter. In addition, the actual rates of transformation between molecules and bound pairs have to be faster than the rates involving exchange between free and bound charges.

The double-equilibrium with charge generation $(00) \rightleftharpoons (+-) \rightleftharpoons (+) + (-)$ studied in the main text can be described by the following set of equations:
\begin{eqnarray}
n_\mathrm{b}/2 &=&  K_e n_e^2 \\
(n_\mathrm{f}/2)^2 &=&  K_\mathrm{c} n_\mathrm{b}/2 \\
n_e + n_\mathrm{b} + n_\mathrm{f} &=&  1 \,.
\end{eqnarray}

Application of the external field shifts the concentration quotient $K_\mathrm{c}$ to $K_\mathrm{c}(E) = K_\mathrm{c} \gamma(0)^2 F(\ell_T / \ell_E) $, while the creation constant $K_e$ stays constant. Solving for the in-field concentrations $n'_f$, and $n'_b$ in terms of the zero-field ones, we find:
\begin{eqnarray}
\frac{n'_f}{n_\mathrm{f}} &=&  \frac{\gamma(0)^2F(\ell_T / \ell_E) n_\mathrm{f}}{2n_\mathrm{b}} \\
	&\times& \left[ \sqrt{\left(1+\frac{1-n_\mathrm{b}-n_\mathrm{f}}{\gamma\sqrt{F(\ell_T / \ell_E) }n_\mathrm{f}}\right)^2 + \frac{4n_\mathrm{b}}{\gamma(0)^2F(\ell_T / \ell_E) n_\mathrm{f}^2} } - \left(1+\frac{1-n_\mathrm{b}-n_\mathrm{f}}{\gamma\sqrt{F(\ell_T / \ell_E) }n_\mathrm{f}}\right) \right] \\
\frac{n'_b}{n_\mathrm{b}} &=&  \frac{\gamma(0)^2F(\ell_T / \ell_E) n_\mathrm{f}^2}{4n_\mathrm{b}^2} \\
	&\times& \left[ \sqrt{\left(1+\frac{1-n_\mathrm{b}-n_\mathrm{f}}{\gamma\sqrt{F(\ell_T / \ell_E) }n_\mathrm{f}}\right)^2 + \frac{4n_\mathrm{b}}{\gamma(0)^2F(\ell_T / \ell_E) n_\mathrm{f}^2} } - \left(1+\frac{1-n_\mathrm{b}-n_\mathrm{f}}{\gamma\sqrt{F(\ell_T / \ell_E) }n_\mathrm{f}}\right) \right]^2 \,.
\end{eqnarray}
Expanding the previous result ($n_e \gg n_\mathrm{f} \gg n_\mathrm{b}$) yields:
\begin{eqnarray}
\frac{n'_f}{n_\mathrm{f}} &=&  \gamma\sqrt{F(\ell_T / \ell_E) } + \left(\gamma\sqrt{F(\ell_T / \ell_E) }-\gamma(0)^2F(\ell_T / \ell_E) \right)n_\mathrm{f} + \mathcal{O}\left(n_\mathrm{b},\,n_\mathrm{f}^2\right) \\
&=& \gamma\sqrt{F(\ell_T / \ell_E) } + \mathcal{O}\left(n_\mathrm{b},\,n_\mathrm{f}\right)\\
\frac{n'_b}{n_\mathrm{b}} &=&  1 + 2\left(1-\gamma\sqrt{F(\ell_T / \ell_E) }\right)n_\mathrm{f} + \mathcal{O}\left(n_\mathrm{b},\,n_\mathrm{f}^2\right) \\
&=& 1 + \mathcal{O}\left(n_\mathrm{b},\,n_\mathrm{f}\right)
\end{eqnarray}

The fraction of bound pairs stays constant up to a linear order in $n_\mathrm{b}$, which is of the order $10^{-5}$ at $\{\mu^* = -1.45 , T^* = 0.140\}$. 

To sum up, the requirements for the kinetics presented in the main text to hold are that empty sites dominate (weak electrolyte) and that the charge creation kinetics is faster than the bound pair dissociation. This is indeed the case in MC simulations, where charge creation is a single step process, while bound pair dissociation takes multiple steps.

\vspace{0.5cm}
\noindent
{\bf 5. Supplementary Discussion: Mobility -- Metropolis algorithm, Onsager--Fuoss theory}

The Metropolis algorithm yields a relative decrease of mobility with field, resulting from the unity probability of stepping in the field direction and probability $\exp(- E^* / T^*)$ of moving against the field for each proposed step:
\begin{equation}
\frac{\omega_0(E^*)}{\omega_0(0)} = \frac{1 - \exp(- E^* / T^*)}{E^* / T^*},
\end{equation}
where $\omega_0$ is the mobility of an equivalent non-interacting particle. Note, that since the proposed steps are local, it should be possible to rescale the time by the acceptance probability to obtain a close approximation to the physical time \cite{JaubertS}. The mobility would still be field-dependent, as predicted e.g. for monopole hopping in spin ice \cite{CMSDH}.

The full zero-field mobility $\omega(0)$ of the Coulomb gas is further reduced because of the ionic atmosphere slowing down the charge. The zero-field correction is well approximated by Fuoss-Onsager conductivity theory~\cite{FuossOnsagerS} (which fails when the parameter $\ell_T/\ell_\mathrm{D}$ approaches unity, which happens around $T^* = 0.16$). Fuoss-Onsager theory for a binary electrolyte takes the following form:
\begin{eqnarray}
\frac{\omega(0)}{\omega_0(0)} &=& 1 - h\left(\ell_T, \ell_\mathrm{D}\right) = \frac{2 - \sqrt{2}}{3} \frac{\ell_T}{\ell_\mathrm{D}} \nonumber\\
&&- \frac{1}{3}\left(\frac{\ell_T}{\ell_\mathrm{D}}\right)^2 \ln\left(\frac{\ell_T}{\ell_\mathrm{D}}\right)
- \left(\frac{\ell_T}{\ell_\mathrm{D}}\right)^2 N\left(\frac{2\ell_T}{a}\right) + \left(\frac{\ell_T}{\ell_\mathrm{D}}\right)^2 \gamma(0)^2 K_\mathrm{c} \nonumber\\
T_1\left(\frac{2\ell_T}{a}\right) &=& \exp\left(-\frac{2\ell_T}{a}\right) \left(1 + \frac{2\ell_T}{a} + \frac{1}{2}\left(\frac{2\ell_T}{a}\right)^2 \right) \nonumber\\
N\left(\frac{2\ell_T}{a}\right) &=& 1.4985 + \frac{0.2071 T_1(2l_T/a) - 0.03066}{1 - T_1(2l_T/a)} \nonumber\\
K_\mathrm{c}\left(\frac{2\ell_T}{a}\right) &=& \frac{1}{3} \left(\mathrm{Ei}\left(\frac{2\ell_T}{a}\right) - \frac{a}{2\ell_T} \left(1 + \frac{a}{2\ell_T}\right) \exp\left(\frac{2\ell_T}{a}\right) \right) 
\,.
\end{eqnarray}

To describe the field dependence of this correction, we use Wilson's calculation of the first Wien effect (for review see Eckstrom and Schmelzer~\cite{EckstromSchmelzer}), which is known to give the field evolution of the term linear in $\ell_T/\ell_\mathrm{D}$ from the previous equation. We assume that the other terms decay in the same or similar fashion; an assumption that appears {\it a posteriori} to be a good approximation. We obtain
\begin{eqnarray}
\frac{\omega(E^*)}{\omega_0(E^*)} &=& 1 - h\left(\ell_T, \ell_\mathrm{D}\right) g\left(\ell_\mathrm{D} / \ell_E \right), \mathrm{where} \nonumber\\
g(x) &=& \frac{3}{(4-2\sqrt{2})x^3}\left[ x\sqrt{1+x^2} - \arctan\left(\frac{x}{\sqrt{1+x^2}}\right) - \sqrt{2}x + \arctan\left(\sqrt{2}x\right) \right] \,,
\end{eqnarray}
and $g(x)$ is scaled to decrease from unity in zero field to zero in infinite field.

The relative change of the total mobility is thus given by:
\begin{equation}
\frac{\omega(E^*)}{\omega(0)} = \frac{1 - h\left(\ell_T, \ell_\mathrm{D}\right) g\left(\ell_\mathrm{D} / \ell_E \right)}{1 - h\left(\ell_T, \ell_\mathrm{D}\right)} \frac{1 - \exp(- E^* / T^*)}{E^* / T^*} \,.
\end{equation}

Note that in all calculations we only use terms from the relaxation field and exclude terms related to electrophoresis which does not occur because there is no hydrodynamics in our simulations, which have only stochastic dynamics.

\vspace{0.5cm}
\noindent
{\bf 6. Supplementary Discussion: Finite-size effects}

The CPU time scales with the number of sites $N \propto L^3$, the memory with $N^2 \propto L^6$, since we save the pre-computed Ewald summation for every pair of sites. Saving the potentials allows us to reduce the multiplicative constant of the $L^3$ CPU time dependence. Because the system size is limited, we need to analyse the finite-size effects influencing the simulations results.

Zero-field concentration of charges shows large finite-size effects if $l_D > L/2$. The concentration decreases with system size because the volume accessible is too small for a pair of particles to become unbound. For $\mu^* = -1.45$ and $L = 24$, the simulation temperature must stay above $T^* = 0.130$. Figures in the main text are at $T^* \gtrsim 0.140$, which is sufficiently far above the limit because concentration grows exponentially with temperature.

A different kind of finite size effects appears at high fields. The screening cloud becomes elongated and charges diffuse fast in the field direction. Due to periodic boundary conditions, the charges eventually start to interact with a copy of their screening cloud that has wrapped around the system -- a partial revival of screening, which decreases $\gamma(E)$, leading to the observed charge concentration exceeding that in thermodynamic limit. We confirmed the process by inspecting the pair correlation functions. For $\mu^* = -1.45$, $L=24$, and $T^* = 0.140$ these effects set in above $E^* = 0.08$.

In many relevant materials, it would be interesting to study quasi-two-dimensional geometries of samples. In that case, the zero-field concentration of charges can be modified severly by the first of the above-mentioned finite size effects and a significant unit-cell polarization can be induced by the second. For boundary conditions of finite dielectric constant, the Ewald summation must then be treated carefully~\cite{YehBerkowitz}.

\newpage
\vspace{0.5cm}
\noindent
{\bf 7. Supplementary Discussion: Approximation of a Real Systems as a Lattice Electrolyte -- the Case of Water Ice}

How can we approximate a real material system -- in all its complexity -- by an idealised lattice electrolyte of the type studied in the main text? Renormalisation group theory provides a general theoretical framework and mathematically rigorous formalism by which to answer such a question, but there are few real systems of interest that are sufficiently simple to allow such an approach. In most practical cases, one would require a controlled series of approximations to effect the passage between the realistic microscopic model and the idealised lattice electrolyte. To illustrate how this can be achieved we consider one specific case, -- that of water ice. This example has the added bonus that the occurrence of Wien effect of water ice appears to remain an open question, despite many years of research. Hence the outline given here recommends a long term programme for how to solve this vexing problem. 

Water ice~\cite{WhitworthPetrenko,Hobbs} is an important real substance and also provides a paradigm for systems with macroscopic degeneracy, having lent its name, for example, to the concept of `spin ice' in magnetism~\cite{BramwellGingras}. Highly realistic models of water ice have been investigated by numerical methods \cite{Ice_numerics}, particularly with regard to surface properties. The bulk properties of water ice are more elusive to simulation as they rely on a small concentration of defects, and the bulk physics of water ice is necessarily based on a number of simplified models of point defects derived from the basic ice model of Bernal--Fowler~\cite{BernalFowler} and Pauling~\cite{Pauling}, and the non-equilibrium thermodynamic treatment of Jaccard~\cite{Jaccard}. 

In the crystal structure of ordinary water ice (Ih) the water ions occupy a tetrahedral lattice with hexagonal space symmetry, the hexagonal relative of the diamond lattice considered here. In the ground state one hydrogen ion or proton occupies each line of contact between oxide ions, but is shifted away from the mid-point of the line so that two protons lie close to, and two lie further away from each oxide. This `ice rule' ensures that the crystal structure is composed of covalently bonded water molecules ${\rm H_2O}$ connected by hydrogen bonds. However, as recognised by Pauling~\cite{Pauling} the ice rule does not order the proton subsystem, so water ice has a zero point entropy which may be measured in experiment~\cite{Giauque}. 

There are four basic types of electrical point defect in the water ice structure~\cite{Jaccard}: the ionic defects ${\rm H_3O^+}$ and ${\rm OH^-}$ and so called D and L defects in which oxide-oxide contacts are occupied by two or zero protons respectively. Without the presence of D/L defects the drift of ionic defects in water ice in response to an applied electric field would form long chains of polarised water molecules in a manner akin to the celebrated `Grotthus mechanism' of conduction. However, the ensuing loss of entropy from the ground state would then provide a thermodynamic force to oppose and eventually extinguish the direct current (such a scenario occurs for magnetic monopoles in spin ice~\cite{RyzhkinS}). The D/L defects play the crucial role of relaxing the polarisation strings and thus allowing the passage of a direct proton current through the boundaries of the sample. 

{\it A priori} there is no reason to expect that transport of ionic defects in water ice should occur by quasi-classical hopping diffusion, and indeed there is experimental evidence of proton tunnelling~\cite{BoveKlotz}. However Chen {\it et al.}~\cite{Chen} studied a realistic tight binding model of proton dynamics in the ice structure and showed how proton tunnelling in the disordered potential landscape of water ice can lead to effective classical diffusion on mesoscopic scales. This justifies the approximation of proton dynamics in water ice by classical hopping diffusion. 

Ryzhkin~\cite{Ryzhkin_pseudospin} proposed a pseudospin hamiltonian for water ice, which is equivalent to the so-called `dipolar spin ice' hamiltonian for spin ice materials~\cite{BramwellGingras}. Applying the transformation of this Hamiltonian described in Ref.~\cite{CMSS}, we can reduce water ice to a lattice electrolyte of Coulombically interacting ${\rm H_3O^+}$ and ${\rm OH^-}$ ions, situated on the oxide lattice and connected by polarisation strings (`Dirac strings'). The effect of D/L defects may then be accounted for by simply neglecting the strings, in which case we arrive at a lattice electrolyte model, equivalent, in thermodynamic terms, to the one studied here (apart from the change in global symmetry from cubic to hexagonal).  Putting in realistic numbers we see that water ice may be justifiably approximated by the lattice electrolyte considered here using the parameters $T^* = 0.038$ and $\mu^* = 1.01$ (Fig. 1a, main text). 

The second Wien effect for water ice was considered by Onsager and Dupuis~\cite{OnsagerDupuis}, who calculated an appropriate dielectric constant to use in Onsager's formula. The resulting calculation was found to be consistent with measurements of Eigen and colleagues~\cite{Eigen-Wien} (Supplementary Figure 1). However, while the Wien effect could be of great importance for the electrical response of water ice at high electric fields below the dielectric breakdown of ice (for example for ice crystals in thunder clouds) it has barely been considered since. This may be due in part to the relative ambiguity of the Eigen--Onsager result (Supplementary Figure 1), and in part to confusion concerning early conductivity results that were not corrected for surface effects~\cite{Hobbs}.

\setcounter{figure}{0}
\renewcommand{\figurename}{Supplementary Figure}
\begin{figure}
\begin{center}
\includegraphics[width=0.5\linewidth]{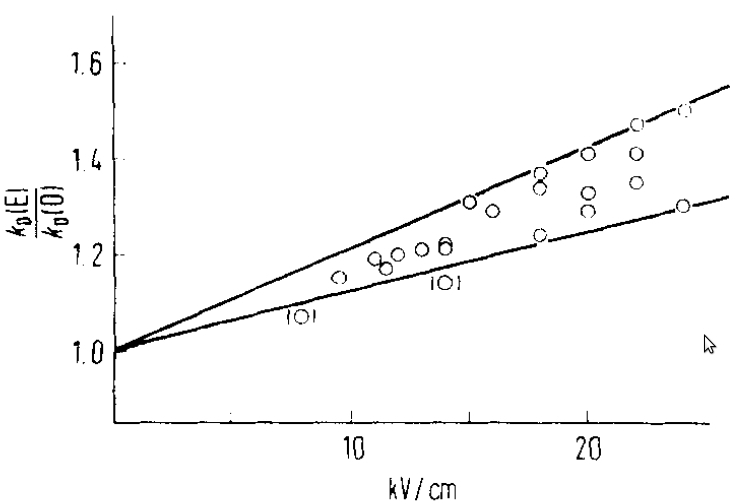}
\end{center}
\caption{
Eigen \& al. measurement of the second Wien effect in water ice\cite{Eigen-Wien}.
}
\label{fig:eigen}
\end{figure}

A more definite theory of the Wien effect on water ice would clearly rely on theoretical estimation of the several factors identified above that were neglected in the original Onsager-Dupuis treatment. Clearly it would then be possible, in a long term programme, to directly re-introduce many of the neglected details of water ice into a numerical simulation, including the local polarisation, D/L defects etc., and to examine their effect on the Onsager theory. Water ice thus affords an example of the controlled passage from realistic microscopic model to idealised lattice electrolyte. 

If one accepts that the various approximations are justified then we immediately arrive at a {\it new prediction} concerning the Wien effect in water ice: that the Wien effect of the conductivity of water ice should be modified by the field-dependent mobility of the lattice electrolyte identified in this paper. As in other cases (see the main text and Supplementary Discussion) this may go some way to regularising the slightly confused situation concerning the comparison of experiment with Onsager's theory in the case of water ice.

\renewcommand\refname{Supplementary Information -- References}

\end{document}